\documentclass{article}
\usepackage[T1]{fontenc} 
\usepackage[utf8]{inputenc} 
\usepackage{ismir,amsmath,cite,url}
\usepackage{graphicx}

\usepackage{color}

\usepackage[firstpage]{draftwatermark}
\definecolor{lightgray}{rgb}{0.9,0.9,0.9}
\definecolor{darkgray}{rgb}{0.4,0.4,0.4}
\SetWatermarkFontSize{12pt}
\SetWatermarkScale{1.1}
\SetWatermarkAngle{90}
\SetWatermarkHorCenter{202mm}
\SetWatermarkVerCenter{170mm}
\SetWatermarkColor{darkgray}
\SetWatermarkText{Late-Breaking / Demo Session Extended Abstract, ISMIR 2024 Conference}

\def\lbd{}

\usepackage{lineno}

\title{HARP 2.0: Expanding Hosted, Asynchronous, Remote Processing for Deep Learning in the DAW}

\multauthor
{Christodoulos Benetatos$^1$* \hspace{0.5cm} Frank Cwitkowitz$^1$* \hspace{0.5cm} Nathan Pruyne$^2$ } { \bfseries{Hugo Flores Garcia$^2$\hspace{0.5cm} Patrick O'Reilly $^2$ \hspace{0.5cm} Zhiyao Duan$^1$ \hspace{0.5cm} Bryan Pardo$^2$}\\
  $^1$ Department of Electrical \& Computer Engineering, University of Rochester, USA\\
$^2$ Department of Computer Science, Northwestern University, USA\\
{\tt\small patrick.oreilly2024@u.northwestern.edu}
\thanks{*Authors contributed equally.}
}

\def\authorname{C. Benetatos and F. Cwitkowitz and N. Pruyne and H.F. Garcia and P. O'Reilly and Z. Duan and B. Pardo}

\begin{document}

\maketitle

\begin{abstract}

HARP 2.0 brings deep learning models to digital audio workstation (DAW) software through \textbf{h}osted, \textbf{a}synchronous, \textbf{r}emote \textbf{p}rocessing, allowing users to route audio from a plug-in interface through any compatible Gradio endpoint to perform arbitrary transformations. HARP renders endpoint-defined controls and processed audio in-plugin, meaning users can explore a variety of cutting-edge deep learning models without ever leaving the DAW. In the 2.0 release we introduce support for MIDI-based models and audio/MIDI labeling models, provide a streamlined \texttt{pyharp} Python API for model developers, and implement numerous interface and stability improvements. Through this work, we hope to bridge the gap between model developers and creatives, improving access to deep learning models by seamlessly integrating them into DAW workflows.

\end{abstract}

\section{Introduction}\label{sec:introduction}

Deep learning models for audio have become valuable tools for creative expression, facilitating tasks from sound synthesis \cite{edmsound} to music transcription \cite{scoringintervals} to remixing \cite{vampnet}. However, these tools are often released as code repositories or are developed as standalone software products. In contrast, many audio content creators work within digital audio workstation (DAW) software that hosts multiple ``plug-in" tools in a unified environment, supporting complex and idiomatic editing workflows. 

HARP \cite{harp} helps to bridge this gap between model developers and users by linking developer-friendly environments (Python-based Gradio \cite{gradio} applications) with user-friendly environments (any popular DAW with support for external sample editing). HARP is a standalone sample-editing application that can be run as a DAW plug-in, allowing users to route audio through any compatible Gradio endpoint without leaving the DAW. Correspondingly, model developers can wrap any standard audio-producing Gradio endpoint with components from the lightweight \texttt{pyharp} API, allowing existing applications to serve audio to HARP with only a few additional lines of code. Model developers can define intuitive interfaces using built-in controls, which are rendered in-plugin once a user selects the corresponding endpoint.

\begin{figure}
    \centering
    \includegraphics[width=0.9\linewidth]{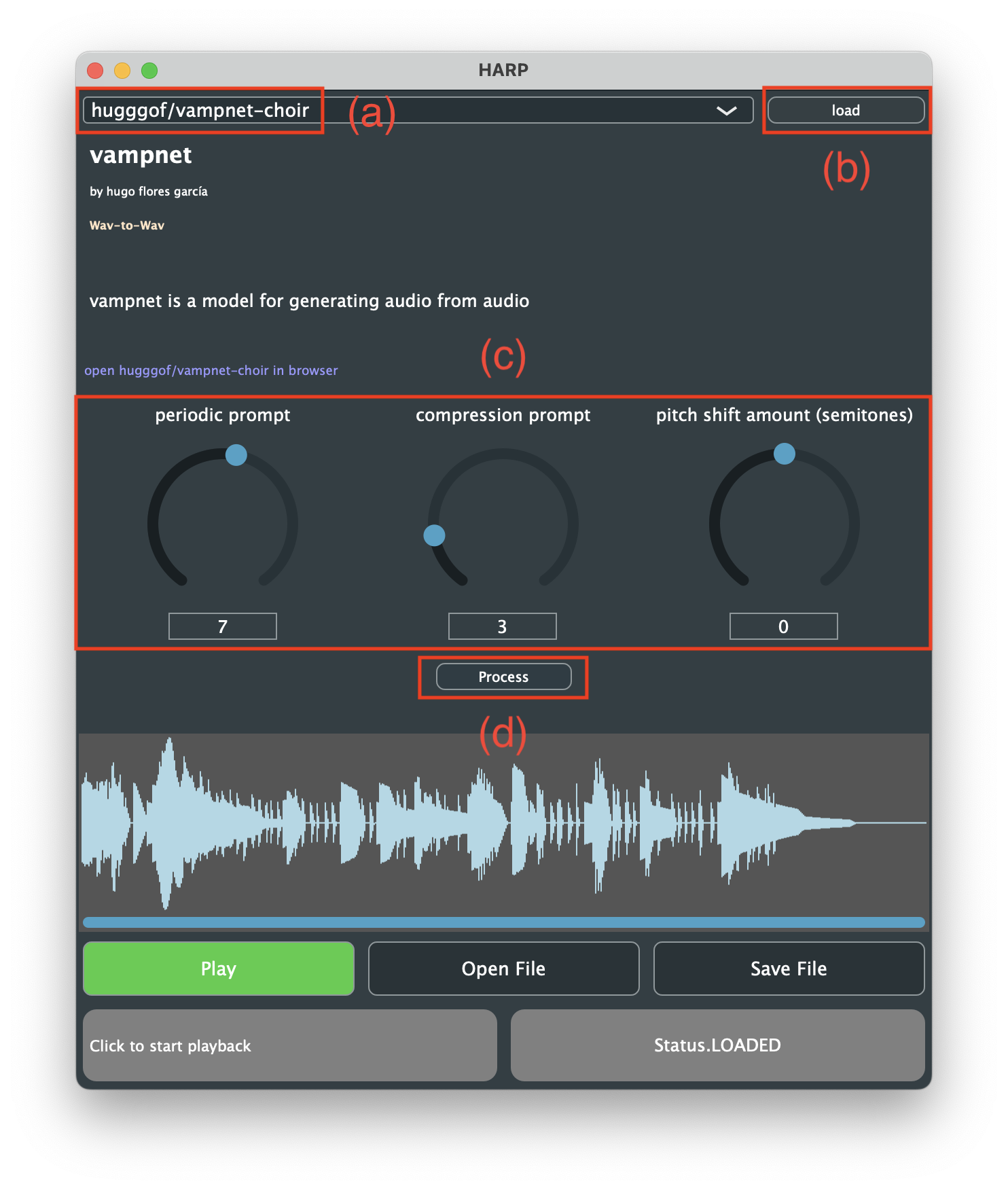}
    \caption{The HARP 2.0 interface. Users can enter a path to a compatible Gradio endpoint or select from a menu of default endpoints (a); upon loading (b), the endpoint's interface (c) is displayed to the user, who can tweak controls before processing (d). Outputs are then rendered for playback and visualization.}
    \label{fig:interface}
\end{figure}

In this demo, we present the 2.0 release of HARP, a comprehensive update including (1) support for MIDI-based deep learning models, which consume and/or produce MIDI files rather than raw audio, (2) support for displaying time-stamped output labels generated from audio or MIDI models and (3) numerous interface and stability improvements, including a complete overhaul of the application format and Windows/Linux support.

We open-source all code\footnote{\url{https://github.com/TEAMuP-dev/HARP}} for HARP 2.0, including example scripts for running audio-to-audio, MIDI-to-MIDI, and labeling models.

\begin{figure}
    \centering
    (a) \includegraphics[width=0.92\linewidth]{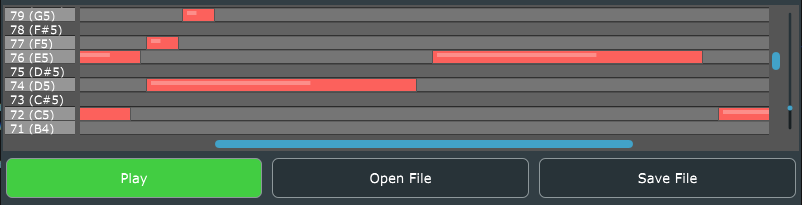}
    (b) \includegraphics[width=0.92\linewidth]{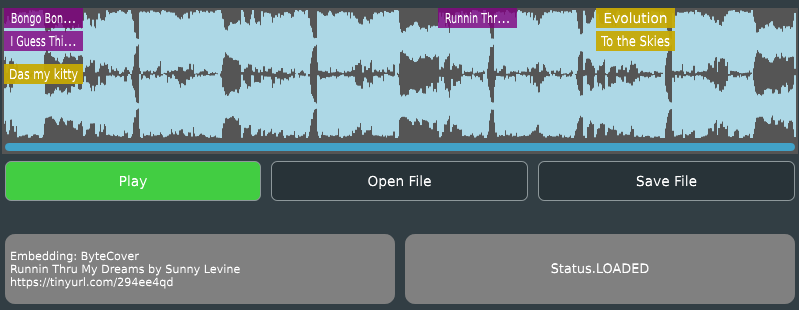}
    \caption{(a) Piano roll for display and playback of MIDI files in HARP.
    (b) Audio-to-labels in HARP. Labels are displayed in multiple colors to denote audio similarity as detected by different embeddings. Details about the label appear in the bottom left `info' box.}
    \label{fig:midi_and_labels}
\end{figure}

\section{MIDI Support}

Digital musicians often compose and edit music in the symbolic MIDI format\footnote{\url{https://en.wikipedia.org/wiki/MIDI}} before rendering audio with interchangeable real-time software synthesizers. A number of deep learning models have been proposed to predict and manipulate MIDI representations \cite{bachduet}, allowing for creative interactions in the symbolic domain. In order to support such symbolic interactions, we extend HARP to allow MIDI input and output, incorporate a standard piano roll display for visualization, and facilitate MIDI playback. An example of MIDI functionality is shown in Figure \ref{fig:midi_and_labels}(a).

\section{Labeling Support}

While deep learning models have demonstrated strong performance in audio creation and manipulation tasks, they can also be used to perform automated analysis of existing audio. For example, speech recognition systems can transcribe recordings, while content attribution systems can identify likely influences for song sections by performing similarity search against a database of music embeddings \cite{musicalroots}. To give creatives access to state-of-the-art deep learning models for audio analysis, we extend HARP to support visualization of user-generated output labels on the media display. In \texttt{pyharp}, time-stamped labels can be created dynamically within the main processing function for a model, with optional durations and long descriptions. This enables placing labels at precise horizontal coordinates on the media display. We also allow for control of the vertical placement by specifying the amplitude (for waveform) or pitch (for pianoroll) coordinates for each label, as well as the color of each label. See Figure \ref{fig:midi_and_labels}(b) for an example of labeling functionality for similarity search with musical embeddings.

\section{Interface}

We migrate HARP from an Audio Random Access (ARA)\footnote{\url{https://en.wikipedia.org/wiki/Audio_Random_Access}} plugin to a standalone sample editor application while retaining the core functionality. This circumvents the limited support for ARA across mainstream DAWs while allowing for a similar interaction paradigm, as most DAWs provide support for linking external sample editors as plugins. As part of this migration, we streamline the interface and introduce a number of quality-of-life improvements, such as a drop-down menu for browsing a selection of HARP-compatible Gradio endpoints. Additionally we add new ``info'' and ``status'' view boxes, which offer guidance based on the current UI element being hovered over and display status updates for the Gradio endpoint, respectively. The HARP 2.0 interface is shown in Figure \ref{fig:interface} for a waveform-to-waveform example, and in Figure \ref{fig:midi_and_labels} for MIDI-to-MIDI and audio-to-labels examples.

\section{Usability \& Stability}

To extend the accessibility of HARP, we introduce support for Windows and Linux machines in addition to Macintosh. We also refactor HARP to take advantage of Gradio's support for \texttt{curl} API queries, removing the need for the bundled Gradio Python client, in turn resulting in a simpler, more lightweight installation. Furthurmore, we streamline the \texttt{pyharp} API and add extensive documentation for developing, deploying, and using HARP-compatible Gradio endpoints. We redesign the error handling system, with improved error codes and messages for both developers and end-users. Finally, we add undo and redo functionality to allow users to easily switch between iterations of model usage.

\section{Conclusion}

HARP 2.0 significantly expands the capabilities of HARP, extending support to MIDI and label-based interaction paradigms and improving the overall user experience. Through the development of HARP, we will continue to improve access to cutting-edge deep learning models for music creators. We also hope that this work fosters stronger connections between deep learning researchers and music creators by bridging the gap between model development and deployment environments, allowing for rapid feedback and meaningful dialogue.

\pagebreak

\bibliography{HARP2}

\end{document}